The interstellar chemistry of $C_3H$ and $C_3H_2$ isomers.

*Jean-Christophe Loison[1]*, Marcelino Agúndez[2], Valentine Wakelam[3], Evelyne Roueff[4], Pierre Gratier[3], Núria Marcelino[5], Dianailys Nuñez Reyes[1], José Cernicharo[2], Maryvonne Gerin[6].*

*Corresponding author: jean-christophe.loison@u-bordeaux.fr

*[1] Institut des Sciences Moléculaires (ISM), CNRS, Univ. Bordeaux, 351 cours de la Libération, 33400, Talence, France*
*[2] Instituto de Ciencia de Materiales de Madrid, CSIC, C\ Sor Juana Inés de la Cruz 3, 28049 Cantoblanco, Spain*
*[3] Laboratoire d'astrophysique de Bordeaux, Univ. Bordeaux, CNRS, B18N, allée Geoffroy Saint-Hilaire, 33615 Pessac, France.*
*[4] LERMA, Observatoire de Paris, PSL Research University, CNRS, Sorbonne Universités, UPMC Univ. Paris 06, F-92190 Meudon, France*
*[5] INAF, Osservatorio di Radioastronomia, via P. Gobetti 101, 40129 Bologna, Italy*
*[6] LERMA, Observatoire de Paris, PSL Research University, CNRS, Sorbonne Universités, UPMC Univ. Paris 06, Ecole Normale Supérieure, F-75005 Paris, France*

We report the detection of linear and cyclic isomers of $C_3H$ and $C_3H_2$ towards various starless cores and review the corresponding chemical pathways involving neutral ($C_3H_x$ with x=1,2) and ionic ($C_3H_x^+$ with x = 1,2,3) isomers. We highlight the role of the branching ratio of electronic Dissociative Recombination (DR) reactions of $C_3H_2^+$ and $C_3H_3^+$ isomers showing that the statistical treatment of the relaxation of $C_3H^*$ and $C_3H_2^*$ produced in these DR reactions may explain the relative c,l-$C_3H$ and c,l-$C_3H_2$ abundances. We have also introduced in the model the third isomer of $C_3H_2$ (HCCCH). The observed cyclic-to-linear $C_3H_2$ ratio vary from 110 ± 30 for molecular clouds with a total density around $1\times10^4$ molecules.cm$^{-3}$ to 30 ± 10 for molecular clouds with a total density around $4\times10^5$ molecules.cm$^{-3}$, a trend well reproduced with our updated model. **The higher ratio for low molecular cloud densities is mainly determined by the importance of the H + l-$C_3H_2$ → H + c-$C_3H_2$ and H + t-$C_3H_2$ → H + c-$C_3H_2$ isomerization reactions.**

**Key words: ISM: abundances - ISM: clouds - ISM: molecules - astrochemistry**

## 1. INTRODUCTION

The $C_3H$ and $C_3H_2$ species are observed in two isomeric forms, cyclic and linear, in various interstellar environments, the cyclic isomers being the most stable isomer in both cases. The ring molecule cyclopropynylidyne (c-$C_3H$) has been widely observed in the galaxy (Turner *et al.* 2000, Fossé *et al.* 2001, Pety *et al.* 2012a, Liszt *et al.* 2014) after its discovery by (Yamamoto *et al.* 1987). The linear counterpart, propynylidyne (l- $C_3H$), was discovered in TMC-1 and IRC+10216 by Thaddeus et al. (1985) and is also widely present in the galaxy (Turner et al. 2000, Fossé et al. 2001, Pety et al. 2012a, McGuire *et al.* 2013), associated to the cyclic isomer but in general with a notably smaller abundance. The cyclopropenylidene (c-$C_3H_2$) has been also observed in many environements (Turner et al. 2000, Fossé et al. 2001, Pety et al. 2012a, McGuire et al. 2013, Liszt et al. 2014, Sakai *et al.* 2010) after its discovery by (Thaddeus et al. 1985). The linear isomer, propadienylidene (l-$C_3H_2$), a cumulene carbon chain ($H_2C$=C=C:), was discovered in TMC-1 by (Cernicharo *et al.* 1991) and in other sources (Cernicharo *et al.* 1999, Turner et al. 2000, Fossé et al. 2001, Pety et al. 2012a, Liszt *et al.* 2012) despite the fact that it is always much less abundant than the cyclic isomer. The linear l-$C_3H_2$ isomer has been proposed as a potential carrier of the 4881 and 5450 Å DIBs (Maier *et al.* 2011), this attribution being finally discarded considering the column densities required to reproduce the DIBs, ranging from two to three orders of magnitude higher than the values measured in the diffuse medium at radio frequencies in absorption (Krełowski *et al.* 2011, Liszt et al. 2012).

The presence of these isomers raises various key questions about their chemistry. In particular, as cyclic and linear $C_3H$ and $C_3H_2$ are not well reproduced by usual models (Agúndez & Wakelam 2013, Sipilä *et al.* 2016), are rings and chains formed from the same progenitors? Also, is there a relation between the abundance ratio of the most stable isomer to the other isomers and their energy differences, relation which is believed to follow the "minimum energy principle" (Lattelais *et al.* 2009) despite some counter examples (Loomis *et al.* 2015, Loison *et al.* 2016)? Moreover, can we explain the fact that, as already noted by (Fossé et al. 2001), the abundance ratio of cyclic to linear isomer is not constant across the galaxy? In addition to these questions, we emphasize the specific behavior of $C_3H^+$ which is found only in 2 PDRs (Pety *et al.* 2012b, McGuire *et al.* 2014), and nowhere else despite thorough searches in other interstellar environments (McGuire et al. 2013, McGuire *et al.* 2015). To get a better picture of $C_3H_x$ chemistry, we performed an extensive and systematic review and update of the KIDA chemical network (Wakelam *et al.* 2015b) using the methodology presented in (Loison *et al.* 2014a) and in Appendix A of this paper, including specific theoretical calculations presented in Appendix B. We have in particular introduced the third isomer of $C_3H_2$, (t-$C_3H_2$: H-CCC-H), a quasi linear species (Nguyen *et al.* 2001, Aguilera-Iparraguirre *et al.* 2008), which may play a non-negligible role as it has a different chemistry than the other $C_3H_2$ isomers. The thermochemical properties of the $C_3H_{x=0-3}$ neutral and ionic isomers are summarized in Table 1. It should be noted that the geometry of l-$C_3H_2^+$ corresponds to that of t-$C_3H_2$, i.e., $HCCCH^+$. We did not introduce the cyclic isomer of $C_3H^+$ nor the third isomeric form of $C_3H_2^+$ ($H_2CCC^+$), which are notably less stable than the other isomers and, considering the chemistry, are not supposed to play an important role.

**Table 1**: $C_3H_{x=1-3}$ neutral and ionic isomers characteristics

| species | $\Delta H_f^{298}$ kJ/mol | μ (Debye) |
|---|---|---|
| c-$C_3H$<br>cyclopropynylidyne | 715 ± 8 kJ/mol<br>(Costes *et al.* 2009) | 2.30 (experimental)<br>(Lovas *et al.* 1992) |
| l-$C_3H$<br>propynylidyne | 727 ± 8 kJ/mol<br>(Costes et al. 2009) | 3.55 (theory)<br>(Woon 1995) |
| l-$C_3H^+$ | 1591 kJ/mol<br>(Costes et al. 2009, Wang *et al.* 2007) | 3.06 (theory)<br>(Huang *et al.* 2013) |
| c-$C_3H^+$ | 1661 kJ/mol<br>(Costes et al. 2009, Wang et al. 2007) | 1.49 (theory)<br>(This work) |
| c-$C_3H_2$<br>cyclopropenylidene | 497 ± 4 kJ/mol<br>(Vazquez *et al.* 2009) | 3.27 (experimental)<br>(Lovas et al. 1992) |
| t-$C_3H_2$ (HCCCH)<br>Propynylidene | 543 ± 8 kJ/mol<br>(Aguilera-Iparraguirre et al. 2008) | 0.51<br>(Nguyen et al. 2001) |
| l-$C_3H_2$ ($H_2CCC$)<br>propadienylidene<br>(vinylidencarbene) | 557 ± 4 kJ/mol<br>(Vazquez et al. 2009) | 4.16 (theory)<br>(Wu *et al.* 2010). |
| c-$C_3H_2^+$ | 1382.8 ± 9.2 kJ/mol<br>(C. Lau & Ng 2006) | 1.2 (theory)<br>(this work) |
| l-$C_3H_2^+$ ($HCCCH^+$) | 1396 ± 16 kJ/mol<br>(Prodnuk *et al.* 1990) | 0 (theory)<br>(Wong & Radom 1993, Taatjes *et al.* 2005) |
| l-$C_3H_2^+$,b ($H_2CCC^+$) | 1565 kJ/mol<br>(C. Lau & Ng 2006) | 3.0 (theory)<br>(this work) |
| $C_3H_3$ ($CH_2CCH$)<br>2-propynyl (propargyl) | 352 ± 4 kJ/mol<br>(Vazquez et al. 2009) | 0.14 (theory)<br>(this work) |
| c-$C_3H_3^+$ | 1076 ± 6 kJ/mol<br>(CCSD(T)-F12/aug-cc-pVQZ calculations by comparison with l-$C_3H_3^+$, this work) | 0 (theory)<br>(This work) |
| l-$C_3H_3^+$ | 1191 ± 4 kJ/mol<br>($\Delta H_f^{298}(C_3H_3)$ + IE($C_3H_3$) = 839.5 ± 0.1 kJ/mol (Jacovella *et al.* | 0.524 (theory)<br>(Huang *et al.* 2011) |
| | 2013)) | |

We present our observations in section 2 and our chemical model including various updates in Section 3. Our conclusions are presented in Section 4. This study is the last part of a large review of $C_3H_{x=0-8}O_{y=0-1}$ chemistry in cold interstellar media (Loison et al. 2016, Hickson *et al.* 2016b, Wakelam *et al.* 2015a).

## 2 Observations

We have carried out observations of cyclic and linear $C_3H$ and $C_3H_2$ toward various cold dark clouds. The observations were done with the IRAM 30m telescope using the frequency-switching technique and the EMIR 3 mm receiver connected to a fast Fourier transform spectrometer providing a spectral resolution of 50 kHz. At the IRAM 30m telescope, spectra are calibrated by comparing the sky emissivity with that of hot and cold loads and using the ATM program (Cernicharo 1985, Pardo *et al.* 2001). The uncertainty in the calibration in the 3 mm band is ~10 %. Pointing was checked on nearby planets and quasars. The error in the pointing is 3" at most. Data reduction, carried out using the software CLASS within the package GILDAS[1], consisted in the subtraction of a baseline generated by smoothing the spectral regions free of lines. The observations of TMC-1 and B1-b are part of a 3 mm line survey (Marcelino *et al.* 2009), a good part of which was observed between January and May 2012 (see details in (Cernicharo *et al.* 2012)). In the cases of L483, Lupus-1A, L1495B, L1521F, and Serpens South 1a, the observations were carried out from September to November 2014 in selected frequency ranges across the 3 mm band (Agúndez *et al.* 2015). Examples of the observed spectra are shown in Fig. 1. The line parameters given in Tables 2, 3, and 4 were obtained by Gaussian fitting to the observed line profiles. In some sources, lines were simultaneously fitted to two Gaussian components because this resulted in a better fit. For example, in B1-b there are two velocity components at $V_{LSR}$ +6.5 and +7.4 km $s^{-1}$. In TMC-1 the existence of the two velocity components is less obvious, but still the lines of some molecules can be decomposed into two velocity components at $V_{LSR}$ +5.7 and +6.0 km $s^{-1}$.

[1] See http://www.iram.fr/IRAMFR/GILDAS

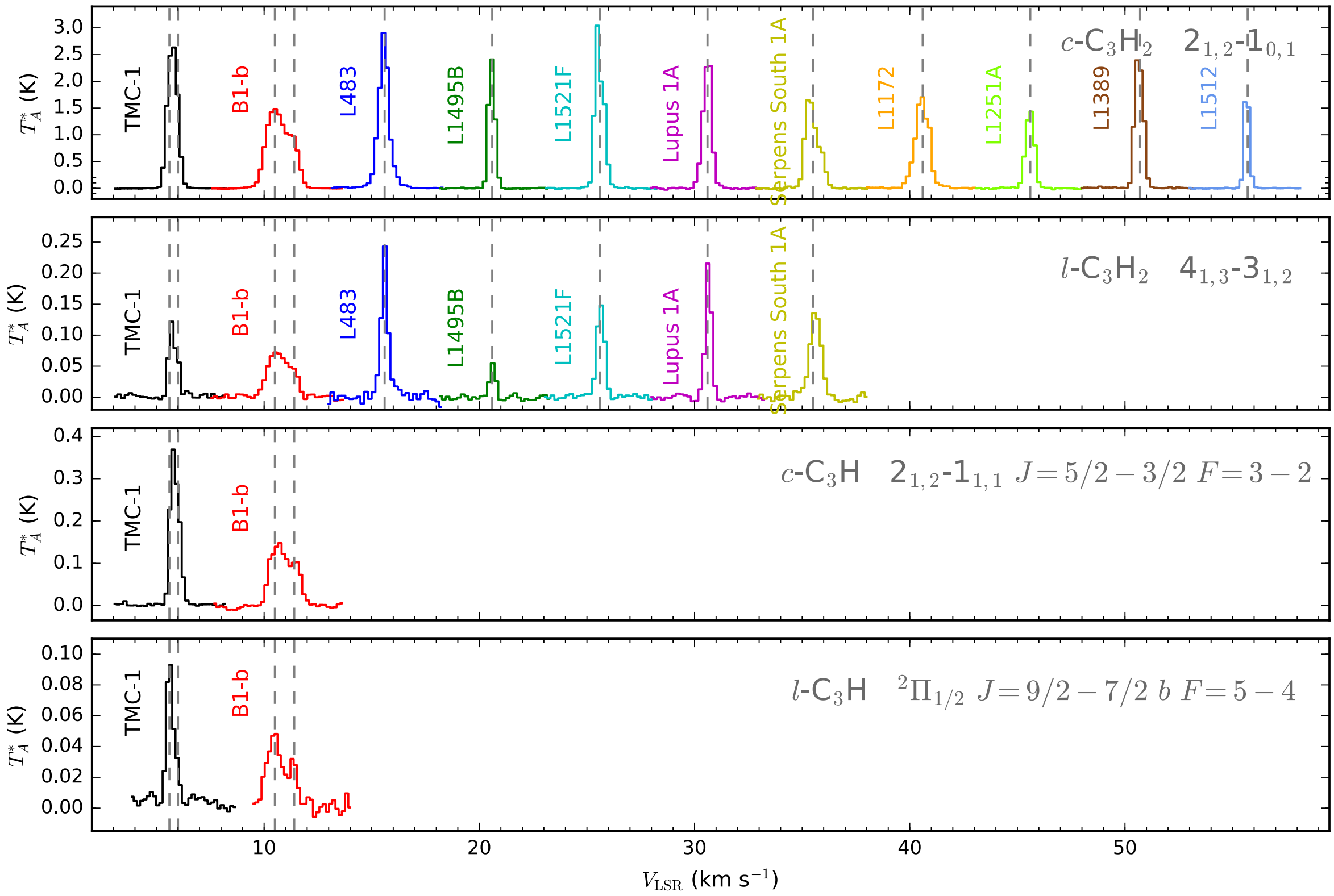

**Figure 1**: Sample of lines of c-$C_3H_2$, l-$C_3H_2$, c-$C_3H$, and l-$C_3H$ observed in the cold dense clouds, where lines of each source have been shifted by fixed amounts along the velocity axis for a better visualization. Dashed vertical lines indicate the position of the $V_{LSR}$ of each source. Note that for TMC-1 and B1-b lines can be fitted to two velocity components. Transition frequencies and quantum numbers of the observed lines are listed in Tables 2, 3 and 4.

**Table 2.** Observed line parameters of $c$-$C_3H_2$ and $l$-$C_3H_2$ in TMC-1 and B1-b.

| Molecule | Transition | Frequency (MHz) | $E_{up}$ (K) | $A_{ul}$ ($s^{-1}$) | HPBW ($''$) | $\frac{B_{eff}}{F_{eff}}$ | $V_{LSR}$ (km $s^{-1}$) | $\Delta v$ (km $s^{-1}$) | $\int T_A^* dv$ (K km $s^{-1}$) |
|---|---|---|---|---|---|---|---|---|---|
| | | | | TMC-1 | | | | | |
| $c$-$C_3H_2$ | $3_{1,2}-3_{0,3}$ (ortho) | 82966.196 | 13.7 | $9.91\times10^{-6}$ | 29.3 | 0.86 | +5.83(2) | 0.46(1) | 0.279(2) |
| | $2_{1,2}-1_{0,1}$ (ortho) | 85338.900 | 4.1 | $2.32\times10^{-5}$ | 28.5 | 0.86 | +5.82(1) | 0.56(1) | 1.659(1) |
| | $4_{3,2}-4_{2,3}$ (ortho) | 85656.415 | 26.7 | $1.52\times10^{-5}$ | 28.4 | 0.86 | +5.81(1) | 0.49(5) | −0.013(1)[a] |
| | $3_{2,2}-3_{1,3}$ (para) | 84727.687 | 16.1 | $1.04\times10^{-5}$ | 28.7 | 0.86 | +5.78(2) | 0.50(2) | 0.078(2) |
| $l$-$C_3H_2$ | $4_{1,3}-3_{1,2}$ (ortho) | 83933.699 | 9.1 | $4.96\times10^{-5}$ | 29.0 | 0.86 | +5.65(2) | 0.24(5) | 0.033(4) |
| | | | | | | | +5.94(2) | 0.27(6) | 0.024(4) |
| | $5_{1,5}-4_{1,4}$ (ortho) | 102992.379 | 13.8 | $9.60\times10^{-5}$ | 23.6 | 0.84 | +5.70(4) | 0.45(3) | 0.022(2) |
| | | | | | | | +6.01(10) | 0.25(10) | 0.002(1) |
| | $5_{1,4}-4_{1,3}$ (ortho) | 104915.583 | 14.1 | $1.02\times10^{-4}$ | 23.2 | 0.84 | +5.65(3) | 0.27(5) | 0.010(2) |
| | | | | | | | +5.93(6) | 0.35(11) | 0.009(2) |
| | $4_{0,4}-3_{0,3}$ (para) | 83165.345 | 10.0 | $5.15\times10^{-5}$ | 29.2 | 0.86 | +5.69(2) | 0.31(4) | 0.018(2) |
| | | | | | | | +5.94(4) | 0.46(8) | 0.015(2) |
| | $5_{0,5}-4_{0,4}$ (para) | 103952.926 | 15.0 | $1.03\times10^{-4}$ | 23.4 | 0.84 | +5.70(5) | 0.36(4) | 0.009(1) |
| | | | | B1-b | | | | | |
| $c$-$C_3H_2$ | $3_{1,2}-3_{0,3}$ (ortho) | 82966.196 | 13.7 | $9.91\times10^{-6}$ | 29.3 | 0.86 | +6.49(1) | 0.88(2) | 0.289(6) |
| | | | | | | | +7.39(1) | 0.58(3) | 0.100(5) |
| | $2_{1,2}-1_{0,1}$ (ortho) | 85338.900 | 4.1 | $2.32\times10^{-5}$ | 28.5 | 0.86 | +6.55(1) | 0.92(1) | 1.492(9) |
| | | | | | | | +7.41(1) | 0.66(1) | 0.597(9) |
| | $4_{3,2}-4_{2,3}$ (ortho) | 85656.415 | 26.7 | $1.52\times10^{-5}$ | 28.4 | 0.86 | +6.54(2) | 0.67(5) | 0.020(1) |
| | | | | | | | +7.45(7) | 0.30(14) | 0.002(1) |
| | $3_{2,2}-3_{1,3}$ (para) | 84727.687 | 16.1 | $1.04\times10^{-5}$ | 28.7 | 0.86 | +6.50(2) | 0.86(3) | 0.087(3) |
| | | | | | | | +7.38(3) | 0.55(5) | 0.025(3) |
| $l$-$C_3H_2$ | $4_{1,3}-3_{1,2}$ (ortho) | 83933.699 | 9.1 | $4.96\times10^{-5}$ | 29.0 | 0.86 | +6.56(4) | 0.95(7) | 0.075(5) |
| | $5_{1,5}-4_{1,4}$ (ortho) | 102992.379 | 13.8 | $9.60\times10^{-5}$ | 23.6 | 0.84 | +6.54(2) | 0.84(5) | 0.054(3) |
| | | | | | | | +7.39(2) | 0.41(5) | 0.017(2) |
| | $5_{1,4}-4_{1,3}$ (ortho) | 104915.583 | 14.1 | $1.02\times10^{-4}$ | 23.2 | 0.84 | +6.53(2) | 0.80(4) | 0.046(2) |
| | | | | | | | +7.47(2) | 0.55(5) | 0.023(2) |
| | $4_{0,4}-3_{0,3}$ (para) | 83165.345 | 10.0 | $5.15\times10^{-5}$ | 29.2 | 0.86 | +6.54(3) | 0.92(9) | 0.043(2) |
| | $5_{0,5}-4_{0,4}$ (para) | 103952.926 | 15.0 | $1.03\times10^{-4}$ | 23.4 | 0.84 | +6.54(3) | 0.89(6) | 0.034(2) |
| | | | | | | | +7.42(4) | 0.50(7) | 0.011(2) |

Numbers in parentheses are $1\sigma$ uncertainties in units of the last digits. Antenna temperature ($T_A^*$) can be converted to main beam brightness temperature ($T_{MB}$) by dividing by ($B_{eff}/F_{eff}$).

[a] Line in absorption against cosmic microwave background.

**Table 3.** Observed line parameters of $c$-$C_3H_2$ and $l$-$C_3H_2$ in various cold dense clouds.

| Molecule | Transition | Frequency (MHz) | $E_{up}$ (K) | $A_{ul}$ ($s^{-1}$) | HPBW ($''$) | $\frac{B_{eff}}{F_{eff}}$ | $V_{LSR}$ (km $s^{-1}$) | $\Delta v$ (km $s^{-1}$) | $\int T_A^* dv$ (K km $s^{-1}$) |
|---|---|---|---|---|---|---|---|---|---|
| L483 | | | | | | | | | |
| $c$-$C_3H_2$ | $2_{1,2}-1_{0,1}$ (ortho) | 85338.900 | 4.1 | $2.32\times10^{-5}$ | 28.5 | 0.86 | +5.26(1) | 0.55(1) | 1.637(9) |
| | $4_{3,2}-4_{2,3}$ (ortho) | 85656.415 | 26.7 | $1.52\times10^{-5}$ | 28.4 | 0.86 | +5.28(2) | 0.58(6) | 0.069(5) |
| | $3_{2,2}-3_{1,3}$ (para) | 84727.687 | 16.1 | $1.04\times10^{-5}$ | 28.7 | 0.86 | +5.29(1) | 0.39(1) | 0.165(2) |
| $l$-$C_3H_2$ | $4_{1,3}-3_{1,2}$ (ortho) | 83933.699 | 9.1 | $4.96\times10^{-5}$ | 29.0 | 0.86 | +5.30(1) | 0.36(1) | 0.091(3) |
| | $5_{1,5}-4_{1,4}$ (ortho) | 102992.379 | 13.8 | $9.60\times10^{-5}$ | 23.6 | 0.84 | +5.25(1) | 0.35(2) | 0.055(2) |
| | $4_{0,4}-3_{0,3}$ (para) | 83165.345 | 10.0 | $5.15\times10^{-5}$ | 29.2 | 0.86 | +5.30(1) | 0.34(2) | 0.054(2) |
| Lupus-1A | | | | | | | | | |
| $c$-$C_3H_2$ | $2_{1,2}-1_{0,1}$ (ortho) | 85338.900 | 4.1 | $2.32\times10^{-5}$ | 28.5 | 0.86 | +5.00(1) | 0.47(1) | 1.259(6) |
| | $3_{2,2}-3_{1,3}$ (para) | 84727.687 | 16.1 | $1.04\times10^{-5}$ | 28.7 | 0.86 | +5.06(1) | 0.36(1) | 0.112(2) |
| $l$-$C_3H_2$ | $4_{1,3}-3_{1,2}$ (ortho) | 83933.699 | 9.1 | $4.96\times10^{-5}$ | 29.0 | 0.86 | +5.05(1) | 0.33(1) | 0.078(1) |
| | $5_{1,5}-4_{1,4}$ (ortho) | 102992.379 | 13.8 | $9.60\times10^{-5}$ | 23.6 | 0.84 | +5.01(1) | 0.33(1) | 0.044(1) |
| | $5_{1,4}-4_{1,3}$ (ortho) | 104915.583 | 14.1 | $8.74\times10^{-5}$ | 23.2 | 0.84 | +5.06(2) | 0.27(4) | 0.039(5) |
| | $4_{0,4}-3_{0,3}$ (para) | 83165.345 | 10.0 | $5.15\times10^{-5}$ | 29.2 | 0.86 | +5.06(1) | 0.35(2) | 0.042(2) |
| | $5_{0,5}-4_{0,4}$ (para) | 103952.926 | 15.0 | $1.03\times10^{-4}$ | 23.4 | 0.84 | +5.07(4) | 0.41(9) | 0.039(7) |
| L1521F | | | | | | | | | |
| $c$-$C_3H_2$ | $2_{1,2}-1_{0,1}$ (ortho) | 85338.900 | 4.1 | $2.32\times10^{-5}$ | 28.5 | 0.86 | +6.36(1) | 0.50(1) | 1.491(9) |
| | $4_{3,2}-4_{2,3}$ (ortho) | 85656.415 | 26.7 | $1.52\times10^{-5}$ | 28.4 | 0.86 | +6.56(3) | 0.32(6) | 0.017(3) |
| | $3_{2,2}-3_{1,3}$ (para) | 84727.687 | 16.1 | $1.04\times10^{-5}$ | 28.7 | 0.86 | +6.43(1) | 0.39(1) | 0.101(2) |
| $l$-$C_3H_2$ | $4_{1,3}-3_{1,2}$ (ortho) | 83933.699 | 9.1 | $4.96\times10^{-5}$ | 29.0 | 0.86 | +6.44(1) | 0.43(1) | 0.071(1) |
| | $5_{1,5}-4_{1,4}$ (ortho) | 102992.379 | 13.8 | $9.60\times10^{-5}$ | 23.6 | 0.84 | +6.41(1) | 0.40(2) | 0.039(1) |
| | $4_{0,4}-3_{0,3}$ (para) | 83165.345 | 10.0 | $5.15\times10^{-5}$ | 29.2 | 0.86 | +6.46(1) | 0.43(2) | 0.039(1) |
| Serpens South 1a | | | | | | | | | |
| $c$-$C_3H_2$ | $2_{1,2}-1_{0,1}$ (ortho) | 85338.900 | 4.1 | $2.32\times10^{-5}$ | 28.5 | 0.86 | +7.33(1) | 0.73(1) | 1.246(12) |
| | $3_{2,2}-3_{1,3}$ (para) | 84727.687 | 16.1 | $1.04\times10^{-5}$ | 28.7 | 0.86 | +7.56(1) | 0.64(1) | 0.146(3) |
| $l$-$C_3H_2$ | $4_{1,3}-3_{1,2}$ (ortho) | 83933.699 | 9.1 | $4.96\times10^{-5}$ | 29.0 | 0.86 | +7.53(1) | 0.61(2) | 0.090(2) |
| | $5_{1,5}-4_{1,4}$ (ortho) | 102992.379 | 13.8 | $9.60\times10^{-5}$ | 23.6 | 0.84 | +7.51(1) | 0.59(3) | 0.050(2) |
| | $4_{0,4}-3_{0,3}$ (para) | 83165.345 | 10.0 | $5.15\times10^{-5}$ | 29.2 | 0.86 | +7.57(2) | 0.71(4) | 0.054(2) |
| L1495B | | | | | | | | | |
| $c$-$C_3H_2$ | $2_{1,2}-1_{0,1}$ (ortho) | 85338.900 | 4.1 | $2.32\times10^{-5}$ | 28.5 | 0.86 | +7.57(1) | 0.37(1) | 0.975(3) |
| | $3_{2,2}-3_{1,3}$ (para) | 84727.687 | 16.1 | $1.04\times10^{-5}$ | 28.7 | 0.86 | +7.63(1) | 0.29(1) | 0.033(1) |
| $l$-$C_3H_2$ | $4_{1,3}-3_{1,2}$ (ortho) | 83933.699 | 9.1 | $4.96\times10^{-5}$ | 29.0 | 0.86 | +7.62(1) | 0.32(2) | 0.019(1) |
| | $5_{1,5}-4_{1,4}$ (ortho) | 102992.379 | 13.8 | $9.60\times10^{-5}$ | 23.6 | 0.84 | +7.60(2) | 0.36(6) | 0.010(1) |
| | $4_{0,4}-3_{0,3}$ (para) | 83165.345 | 10.0 | $5.15\times10^{-5}$ | 29.2 | 0.86 | +7.68(1) | 0.18(9) | 0.007(1) |
| L1172 | | | | | | | | | |
| $c$-$C_3H_2$ | $2_{1,2}-1_{0,1}$ (ortho) | 85338.900 | 4.1 | $2.32\times10^{-5}$ | 28.5 | 0.86 | +2.68(1) | 0.75(1) | 1.358(6) |
| L1251A | | | | | | | | | |
| $c$-$C_3H_2$ | $2_{1,2}-1_{0,1}$ (ortho) | 85338.900 | 4.1 | $2.32\times10^{-5}$ | 28.5 | 0.86 | −4.02(1) | 0.47(1) | 0.753(7) |
| L1389 | | | | | | | | | |
| $c$-$C_3H_2$ | $2_{1,2}-1_{0,1}$ (ortho) | 85338.900 | 4.1 | $2.32\times10^{-5}$ | 28.5 | 0.86 | −4.76(1) | 0.47(1) | 1.280(12) |
| L1512 | | | | | | | | | |
| $c$-$C_3H_2$ | $2_{1,2}-1_{0,1}$ (ortho) | 85338.900 | 4.1 | $2.32\times10^{-5}$ | 28.5 | 0.86 | +7.05(1) | 0.28(1) | 0.601(6) |

Numbers in parentheses are $1\sigma$ uncertainties in units of the last digits. Antenna temperature ($T_A^*$) can be converted to main beam brightness temperature ($T_{MB}$) by dividing by ($B_{eff}/F_{eff}$).

**Table 4.** Observed line parameters of $c$-$C_3H$ and $l$-$C_3H$ in TMC-1 and B1-b.

| Molecule | Transition | Frequency (MHz) | $E_{up}$ (K) | $A_{ul}$ ($s^{-1}$) | HPBW ($''$) | $\frac{B_{eff}}{F_{eff}}$ | $V_{LSR}$ (km $s^{-1}$) | $\Delta v$ (km $s^{-1}$) | $\int T_A^* dv$ (K km $s^{-1}$) |
|---|---|---|---|---|---|---|---|---|---|
| | | | | TMC-1 | | | | | |
| $c$-$C_3H$ | $2_{1,2}-1_{1,1}$ $J$=5/2-3/2 $F$=3-2 | 91494.349 | 4.4 | $1.46\times10^{-5}$ | 26.6 | 0.86 | +5.70(6) | 0.27(2) | 0.080(1) |
| | | | | | | | +5.96(16) | 0.41(3) | 0.110(1) |
| | $2_{1,2}-1_{1,1}$ $J$=5/2-3/2 $F$=2-1 | 91497.608 | 4.4 | $1.26\times10^{-5}$ | 26.6 | 0.86 | +5.79(2) | 0.28(2) | 0.057(1) |
| | | | | | | | +6.05(4) | 0.48(5) | 0.063(1) |
| | $2_{1,2}-1_{1,1}$ $J$=5/2-3/2 $F$=2-2 | 91512.969 | 4.4 | $1.61\times10^{-6}$ | 26.6 | 0.86 | +5.74(10) | 0.30(15) | 0.009(1) |
| | | | | | | | +5.95(20) | 0.46(13) | 0.010(1) |
| | $2_{1,2}-1_{1,1}$ $J$=3/2-1/2 $F$=1-1 | 91681.696 | 4.4 | $2.85\times10^{-6}$ | 26.5 | 0.85 | +5.72(4) | 0.44(4) | 0.019(1) |
| | $2_{1,2}-1_{1,1}$ $J$=3/2-1/2 $F$=1-0 | 91692.752 | 4.4 | $8.15\times10^{-6}$ | 26.5 | 0.85 | +5.63(2) | 0.38(4) | 0.033(3) |
| | | | | | | | +5.92(7) | 0.34(10) | 0.019(3) |
| | $2_{1,2}-1_{1,1}$ $J$=3/2-1/2 $F$=2-1 | 91699.471 | 4.4 | $1.26\times10^{-5}$ | 26.5 | 0.85 | +5.75(2) | 0.34(9) | 0.085(2) |
| | | | | | | | +6.05(2) | 0.31(20) | 0.032(2) |
| | $2_{1,2}-1_{1,1}$ $J$=3/2-3/2 $F$=1-1 | 91747.372 | 4.4 | $3.27\times10^{-6}$ | 26.5 | 0.85 | +5.69(4) | 0.43(9) | 0.017(3) |
| | | | | | | | +6.06(15) | 0.68(19) | 0.016(3) |
| | $2_{1,2}-1_{1,1}$ $J$=3/2-3/2 $F$=2-2 | 91780.518 | 4.4 | $2.05\times10^{-6}$ | 26.5 | 0.85 | +5.79(2) | 0.34(5) | 0.018(2) |
| | | | | | | | +6.16(6) | 0.32(10) | 0.005(1) |
| $l$-$C_3H$ | $^2\Pi_{1/2}$ $J$=9/2-7/2 b $F$=5-4 | 97995.166 | 12.5 | $6.12\times10^{-5}$ | 24.8 | 0.85 | +5.59(2) | 0.30(2) | 0.030(3) |
| | | | | | | | +5.89(5) | 0.33(7) | 0.010(3) |
| | $^2\Pi_{1/2}$ $J$=9/2-7/2 b $F$=4-3 | 97995.913 | 12.5 | $5.95\times10^{-5}$ | 24.8 | 0.85 | +5.59(2) | 0.34(2) | 0.028(3) |
| | | | | | | | +5.94(2) | 0.23(7) | 0.005(3) |
| | $^2\Pi_{1/2}$ $J$=9/2-7/2 a $F$=5-4 | 98011.611 | 12.5 | $6.13\times10^{-5}$ | 24.8 | 0.85 | +5.62(3) | 0.36(2) | 0.036(2) |
| | | | | | | | +6.00(6) | 0.25(7) | 0.006(1) |
| | $^2\Pi_{1/2}$ $J$=9/2-7/2 a $F$=4-3 | 98012.524 | 12.5 | $5.96\times10^{-5}$ | 24.8 | 0.85 | +5.54(4) | 0.36(6) | 0.028(2) |
| | | | | | | | +5.90(10) | 0.34(14) | 0.009(1) |
| | | | | B1-b | | | | | |
| $c$-$C_3H$ | $2_{1,2}-1_{1,1}$ $J$=5/2-3/2 $F$=3-2 | 91494.349 | 4.4 | $1.46\times10^{-5}$ | 26.6 | 0.86 | +6.46(2) | 0.88(4) | 0.143(10) |
| | | | | | | | +7.31(2) | 0.60(5) | 0.060(6) |
| | $2_{1,2}-1_{1,1}$ $J$=5/2-3/2 $F$=2-1 | 91497.608 | 4.4 | $1.26\times10^{-5}$ | 26.6 | 0.86 | +6.70(5) | 0.90(10) | 0.104(10) |
| | | | | | | | +7.58(5) | 0.55(8) | 0.033(4) |
| | $2_{1,2}-1_{1,1}$ $J$=5/2-3/2 $F$=2-2 | 91512.969 | 4.4 | $1.61\times10^{-6}$ | 26.6 | 0.86 | +6.79(6) | 0.43(14) | 0.006(2) |
| | $2_{1,2}-1_{1,1}$ $J$=3/2-1/2 $F$=1-1 | 91681.696 | 4.4 | $2.85\times10^{-6}$ | 26.5 | 0.85 | +6.60(9) | 0.74(20) | 0.010(2) |
| | $2_{1,2}-1_{1,1}$ $J$=3/2-1/2 $F$=1-0 | 91692.752 | 4.4 | $8.15\times10^{-6}$ | 26.5 | 0.85 | +6.64(5) | 1.25(11) | 0.049(4) |
| | $2_{1,2}-1_{1,1}$ $J$=3/2-1/2 $F$=2-1 | 91699.471 | 4.4 | $1.26\times10^{-5}$ | 26.5 | 0.85 | +6.57(3) | 0.70(5) | 0.076(6) |
| | | | | | | | +7.38(4) | 0.72(8) | 0.048(6) |
| | $2_{1,2}-1_{1,1}$ $J$=3/2-3/2 $F$=1-1 | 91747.372 | 4.4 | $3.27\times10^{-6}$ | 26.5 | 0.85 | +6.49(9) | 0.60(15) | 0.008(2) |
| | $2_{1,2}-1_{1,1}$ $J$=3/2-3/2 $F$=2-2 | 91780.518 | 4.4 | $2.05\times10^{-6}$ | 26.5 | 0.85 | +6.79(9) | 1.03(16) | 0.016(2) |
| $l$-$C_3H$ | $^2\Pi_{1/2}$ $J$=9/2-7/2 b $F$=5-4 | 97995.166 | 12.5 | $6.12\times10^{-5}$ | 24.8 | 0.85 | +6.47(2) | 0.82(5) | 0.040(2) |
| | | | | | | | +7.36(2) | 0.40(6) | 0.013(2) |
| | $^2\Pi_{1/2}$ $J$=9/2-7/2 b $F$=4-3 | 97995.913 | 12.5 | $5.95\times10^{-5}$ | 24.8 | 0.85 | +6.49(3) | 0.84(7) | 0.030(2) |
| | | | | | | | +7.44(6) | 0.43(13) | 0.006(2) |
| | $^2\Pi_{1/2}$ $J$=9/2-7/2 a $F$=5-4 | 98011.611 | 12.5 | $6.13\times10^{-5}$ | 24.8 | 0.85 | +6.51(2) | 0.73(5) | 0.036(2) |
| | | | | | | | +7.36(4) | 0.54(9) | 0.013(2) |
| | $^2\Pi_{1/2}$ $J$=9/2-7/2 a $F$=4-3 | 98012.524 | 12.5 | $5.96\times10^{-5}$ | 24.8 | 0.85 | +6.40(3) | 0.69(7) | 0.027(2) |
| | | | | | | | +7.24(8) | 0.47(10) | 0.008(2) |

Numbers in parentheses are $1\sigma$ uncertainties in units of the last digits. Antenna temperature ($T_A^*$) can be converted to main beam brightness temperature ($T_{MB}$) by dividing by ($B_{eff}/F_{eff}$).

## 2.1 Observed abundances of c/l-$C_3H_2$ and c/l-$C_3H$

From the observed line intensities, we determined column densities of c/l-$C_3H_2$ and c/l-$C_3H$ in the different sources, averaged over the beam of the IRAM 30-m telescope, which, at the observed frequencies, ranges from 23” to 29”. These column densities are given in Table 5 as well as the physical conditions corresponding to the various sources.

**Table 5.** Physical properties, and derived column densities and cyclic-to-linear column density ratios.

| Source | $T_k$ (K) | $n(H_2)$ ($cm^{-3}$) | $N(c\text{-}C_3H)^a$ ($cm^{-2}$) | $N(l\text{-}C_3H)^a$ ($cm^{-2}$) | $N(c\text{-}C_3H_2)^b$ ($cm^{-2}$) | $N(l\text{-}C_3H_2)^a$ ($cm^{-2}$) | $R_1{}^c$ | $R_2{}^c$ |
|---|---|---|---|---|---|---|---|---|
| TMC-1 | 10 | $3.0 \times 10^4$ | $6.1 \times 10^{12}$ | $1.1 \times 10^{12}$ | $1.2 \times 10^{14}$ | $1.8 \times 10^{12}$ | 5.5 | 67 |
| B1-b | 12 | $4.0 \times 10^5$ | $6.2 \times 10^{12}$ | $1.2 \times 10^{12}$ | $1.8 \times 10^{13}$ | $6 \times 10^{11}$ | 5.2 | 30 |
| L483 | 10 | $3.0 \times 10^5$ | | | $4.0 \times 10^{13}$ | $7 \times 10^{11}$ | | 57 |
| Lupus-1A | 14 | $4.0 \times 10^5$ | | | $1.4 \times 10^{13}$ | $3.5 \times 10^{11}$ | | 40 |
| L1521F | 10 | $4.0 \times 10^5$ | | | $1.5 \times 10^{13}$ | $4 \times 10^{11}$ | | 38 |
| Serpens South 1a | 11 | $4.0 \times 10^5$ | | | $1.4 \times 10^{13}$ | $5 \times 10^{11}$ | | 28 |
| L1495B | 10 | $1.1 \times 10^4$ | | | $2.0 \times 10^{14}$ | $1.8 \times 10^{12}$ | | 111 |
| L1172 | 10 | $7.5 \times 10^4$ | | | $1.4 \times 10^{13}$ | | | |
| L1251A | 10 | $2.1 \times 10^4$ | | | $2.2 \times 10^{13}$ | | | |
| L1389 | 10 | $5.2 \times 10^4$ | | | $2.2 \times 10^{13}$ | | | |
| L1512 | 10 | $2.6 \times 10^4$ | | | $1.7 \times 10^{13}$ | | | |

[a] Estimated error is a factor of 2.
[b] Estimated error is 50 % for the first 7 sources and a factor of 2 for L1172, L1251A, L1389, and L1512.
[c] $R_1 = N(c\text{-}C_3H)/N(l\text{-}C_3H)$ and $R_2 = N(c\text{-}C_3H_2)/N(l\text{-}C_3H_2)$. Estimated uncertainties in $R_1$ and $R_2$ are 30 %.
The error estimates and the assumed parameters are described in the text.

### 2.1.1. Statistical equilibrium: c-$C_3H_2$ and l-$C_3H_2$

In the cases of cyclic and linear $C_3H_2$ we carried out statistical equilibrium calculations using the LVG method and assuming a medium with uniform volume density of $H_2$ and gas kinetic temperature. These two physical parameters were taken from the literature for each source, and, when possible (i.e., when multiple lines of c-$C_3H_2$ or l-$C_3H_2$ were available), additional constraints were provided by the line intensity ratios observed in this study. The adopted values of $T_k$ and $n(H_2)$ are given in Table 5. The gas kinetic temperature in most sources is around 10 K, whereas densities span over almost two orders of magnitude in the range $10^4 - 10^6$ $cm^{-3}$. It is important to note that the column densities of cyclic and linear $C_3H_2$ derived are very sensitive to the adopted volume density (the higher the density the lower the column density needed to reproduce the observed line intensities), although the cyclic-to-linear ratios show little dependence with $n(H_2)$. The ortho and para species of both c-$C_3H_2$ and l-$C_3H_2$ were considered independently in the statistical equilibrium calculations and the statistical ortho-to-para ratio of 3 was assumed for both isomers, which is consistent with the intensity ratios between ortho and para lines observed. To derive the column densities, we carried out various LVG models in which the column density of $C_3H_2$ was varied (for those sources for which multiple lines were available the volume density was also varied) and selected the model that results in the best overall agreement between observed and calculated velocity-integrated intensities. When some of the lines were calculated to be optically thick, as occurs for the $2_{1,2} - 1_{0,1}$ line of c-$C_3H_2$ in most sources, we prioritized to reproduce the intensities of those lines that are optically thin.

For c-$C_3H_2$, level energies were computed from the rotational constants derived by Bogey et al. (1987) and line strengths were computed from the dipole moment of 3.27 debye measured by Lovas et al. (1992), slightly lower than the value of 3.43 debye previously measured by Kanata et al. (1987). Rate coefficients for de-excitation of c-$C_3H_2$ through inelastic collisions with He have been calculated by Avery & Green (1989) including 16 and 17 levels for ortho and para c-$C_3H_2$, respectively, and covering the temperature range 10-30 K, and by Chandra & Kegel (2000) accounting for 47 and 48 levels for ortho and para c-$C_3H_2$, respectively, and covering the temperature range 30-120 K. Unfortunately, Chandra & Kegel (2000) did not extend their calculations down to 10 K. Therefore, here we adopted the values of Avery & Green (1989). We note however that by comparing the rate coefficients calculated in both studies at 30 K the values of Avery & Green (1989) are systematically higher (by up to a factor of 2) than those of Chandra & Kegel (2000), and thus at low densities it may lead to column densities of c-$C_3H_2$ significantly lower than those obtained using the coefficients of Chandra & Kegel (2000). In this study the collision rate coefficients of Avery & Green (1989) at 10 K were adopted, scaled up by a factor of 1.38 to account for $H_2$ instead of He as collider.

In the case of l-$C_3H_2$, level energies were computed from the CDMS archive[2] (Müller *et al.* 2005), which is based on the experimental data of Vrtilek et al. (1990), and a dipole moment of 4.16 debye was adopted, based on the ab initio calculations of Wu et al. (2010). As rate coefficients for de-excitation of l-$C_3H_2$ through inelastic collisions with $H_2$ and He we adopted those calculated for $H_2CO$ by Green (1991).

The density of $H_2$ at the position of the cyanopolyyne peak in TMC-1 has been estimated as $(3 - 8) \times 10^4$ cm$^{-3}$ by Pratap et al. (1997), based on observations of three transitions of $HC_3N$, and around $3 \times 10^4$ cm$^{-3}$ by Lique et al. (2006), based on the observations of multiple transitions of SO. Here we adopt this latter value, which was also adopted by Fossé et al. (2001) in their study of c/l-$C_3H_2$ in TMC-1. The fact that the $4_{3,2} - 4_{2,3}$ transition at 85.6 GHz, with an upper level energy of 26.7 K, is observed in absorption against the cosmic microwave background (see Table 1) indicates that this transition has an excitation temperature below 2.7 K and that c-$C_3H_2$ is not thermalized in TMC-1, which is consistent with the adopted value of $n(H_2)$. From the intensities of the 4 lines of c-$C_3H_2$ observed in TMC-1 we derive a column density of $1.2 \times 10^{14}$ $cm^{-2}$. This value is twice higher than that derived by Fossé et al. (2001) based on the observation of the $1_{1,0} - 1_{0,1}$ transition at 18.3 GHz with the Effelsberg 100-m telescope, probably due to optical depth effects and to the different beam sizes of the Effelsberg 100-m and IRAM 30-m telescopes. Our analysis indicates that the $2_{1,2} - 1_{0,1}$ line at 85.3 GHz is optically thick ($\tau \approx 10$), and thus the derived c-$C_3H_2$ column density relies heavily on the other, optically thinner, observed transitions. The column density of l-$C_3H_2$ derived in TMC-1 is $1.8 \times 10^{12}$ $cm^{-2}$, in agreement with values derived by Cernicharo et al. (1991) and Fossé et al. (2001), and the cyclic-to-linear $C_3H_2$ ratio is 67.

---

[2] See http://www.astro.uni-koeln.de/cdms

The source Barnard 1b is a dense core with a steep density gradient, with values ranging from a few $10^6$ $cm^{-3}$ at the center to a few $10^5$ $cm^{-3}$ at angular scales of the order of the IRAM 30-m beam, according to the modelling of the continuum emission at 350 μm and 1.2 mm by Daniel et al. (2014). Based on this latter study, here we adopt a mean gas kinetic temperature of 12 K and $n(H_2) = 4 \times 10^5$ $cm^{-3}$, physical parameters that result in line intensity ratios in good agreement with the observed ones. The column densities we derive in B1-b are N(c-$C_3H_2$) = $1.8 \times 10^{13}$ $cm^{-2}$ and N(l-$C_3H_2$) = $6 \times 10^{11}$ $cm^{-2}$, which results in a cyclic-to-linear $C_3H_2$ ratio of 30.

The dense core L483 hosts a Class 0 source and shows evidences of infall motions and a density gradient (Tafalla *et al.* 2000). The gas kinetic temperature has been estimated as 10 K (Anglada *et al.* 1997). Densities of $H_2$ around $3 \times 10^4$ $cm^{-3}$ are derived by Anglada et al. (1997), based on $NH_3$ observations, and by Jørgensen et al. (2002), based on the modelling of continuum observations at 450 μm and 850 μm, while Tafalla et al. (2000) derive a mean density of $H_2$ of $3 \times 10^5$ $cm^{-3}$ over a size ≈20" in radius, based on the analysis of $CH_3OH$ emission. Here we adopt this latter value because at a gas kinetic temperature of 10 K, the $4_{3,2} - 4_{2,3}$ transition of c-$C_3H_2$ can only be excited up to the level observed in L483 at densities well above $10^5$ $cm^{-3}$.

In the starless core Lupus-1A, the kinetic temperature has been estimated as 14 ± 2 K (Agúndez et al. 2015), the value adopted here, and as 12.6 ± 1.5 K at a position 2.8' away from Lupus-1A (Sakai *et al.* 2009). The density of $H_2$ has not been accurately estimated but Sakai et al. (2009) point to densities as high as $10^6$ $cm^{-3}$, while the observed intensity ratio between the $2_{1,2} - 1_{0,1}$ and $3_{2,2} - 3_{1,3}$ transitions of c-$C_3H_2$ points as well to densities above $10^5$ $cm^{-3}$. The value of $n(H_2)$ adopted here, $4 \times 10^5$ $cm^{-3}$, results in line intensity ratios for cyclic and linear $C_3H_2$ in reasonable agreement with the observed ones. The column densities derived are N(c-$C_3H_2$) = 1.4 × $10^{13}$ $cm^{-2}$ and N(l-$C_3H_2$) = 3.5 × $10^{11}$ $cm^{-2}$, which results in a cyclic-to-linear $C_3H_2$ ratio of 40.

In L1521F the volume density of $H_2$ has been estimated as 1.1 × $10^6$ $cm^{-3}$, based on observations of the continuum at 1.2 mm, and in the range (1.3 − 4.5) × $10^5$ $cm^{-3}$ from observations of $N_2H^+$ and $N_2D^+$ (Crapsi *et al.* 2005). These authors derive a rotational temperature of 4.8 K for $N_2H^+$, which is consistent with a gas kinetic temperature of 10 K. We find that a density of $H_2$ of 4 × $10^5$ $cm^{-3}$ results in line intensity ratios for cyclic and linear $C_3H_2$ in agreement with the observed ones, in particular as concerns the intensity ratio between the $2_{1,2} - 1_{0,1}$ and $4_{3,2} - 4_{2,3}$ transitions of c-$C_3H_2$. The column densities we derive are N(c-$C_3H_2$) = 1.5 × $10^{13}$ $cm^{-2}$ and N(l-$C_3H_2$) = 4 × $10^{11}$ $cm^{-2}$, which results in a cyclic-to-linear $C_3H_2$ ratio of 38.

In the clump 1a of the Serpens South complex, Friesen et al. (2013) derive a gas kinetic temperature of 11 K through observations of $NH_3$. The density of $H_2$ has not been accurately determined but Friesen et al. (2013) suggest a volume density around $10^4$ $cm^{-3}$ based on timescale arguments. However, on the assumption that the ortho-to-para ratio is 3 for c-$C_3H_2$, the observed intensity ratio between the $2_{1,2} - 1_{0,1}$ and $3_{2,2} - 3_{1,3}$ transitions requires gas densities above $10^5$ $cm^{-3}$. Adopting n($H_2$) = 4 × $10^5$ $cm^{-3}$, the column densities we derive are N(c-$C_3H_2$) = 1.4 × $10^{13}$ $cm^{-2}$ and N(l-$C_3H_2$) = 5 × $10^{11}$ $cm^{-2}$, which results in a cyclic-to-linear $C_3H_2$ ratio of 28.

The physical conditions in the sources L1495B, L1172, L1251A, L1389, and L1512 have been taken from the study by Cordiner et al. (2013). These authors adopt a common gas kinetic temperature of 10 K and derive the volume density from the observation of a couple of lines of $HC_3N$, which are in the range (1 − 10) × $10^4$ $cm^{-3}$. In 4 of these sources, observations are restricted to just one line, the $2_{1,2} - 1_{0,1}$ transition of c-$C_3H_2$, and thus it is not possible to put additional constraints on the volume density. Moreover, according to our statistical equilibrium calculations, this line is optically thick ($\tau$ = 2 − 6) in these sources, which introduces a higher degree of uncertainty in the column densities of c-$C_3H_2$ derived.

The column densities derived for c-$C_3H_2$ and l-$C_3H_2$ have uncertainties coming from the observations (line fitting plus telescope calibration), although the main source of error arises in the conversion of line intensities into column densities, with uncertainties in the kinetic temperature, gas density, collisional rate coefficients, and the probable lack of uniformity of the physical conditions of the cloud within the telescope beam. We estimate that the column densities derived for c-$C_3H_2$ have an uncertainty of 50 % in those sources in which the availability of multiple lines permit to set constraints on the gas density (TMC-1, B1-b, L483,

Lupus-1A, L1521F, Serpens South 1a, L1495B) and an error of a factor of 2 in those sources in which only the $2_{1,2} - 1_{0,1}$ line was observed (L1172, L1251A, L1389, L1512). In the case of l-$C_3H_2$, in those sources in which this isomer was observed, multiple lines were available. However, the adopted collision rate coefficients were not specifically computed for this molecule but for formaldehyde, and thus we estimate uncertainties of a factor of 2 for the column densities of l-$C_3H_2$. Systematic errors in the kinetic temperature and gas density would tend to cancel when computing cyclic-to-linear column density ratios, and thus we estimate errors of 30 % for the cyclic-to-linear $C_3H_2$ ratios derived.

#### 2.2.1 Local thermodynamic equilibrium: c-$C_3H$ and l-$C_3H$

For cyclic and linear $C_3H$ local thermodynamic equilibrium was assumed (e.g. Turner (1991)) due to the lack of adequate collision rate coefficients for these two isomers. The spectroscopy of c-$C_3H$ was taken from JPL database[3] (Pickett *et al.* 1998), which is based on the experimental study by Yamamoto et al. (1994), and the adopted dipole moment is 2.30 debye, the value measured by Lovas et al. (1992). For l-$C_3H$, level energies were taken from the CDMS database (Müller et al. 2005), which is based on laboratory work by Yamamoto et al. (1990) plus some further studies, and we adopted a dipole moment of 3.55 debye based on ab initio calculations by Woon (1995). In this study, observations of c-$C_3H$ and l-$C_3H$ were restricted to various hyperfine components of a single rotational transition in TMC-1 and B1-b. Column densities were derived by assuming a rotational temperature of 5 K for the two isomers in both TMC-1 and B1-b, which is consistent with values from the literature (Mangum & Wootten 1990, Fossé et al. 2001). The column densities derived imply that the cyclic isomer of $C_3H$ is ≈5 times more abundant than the linear one in both TMC-1 and B1-b. We estimate uncertainties of a factor of 2 for the column densities of c-$C_3H$ and l-$C_3H$ in TMC-1 and B1-b, while the cyclic-to-linear $C_3H$ ratios are estimated to have errors of 30 %.

## 3 The chemical model

### 3.1 Model description

To calculate the abundances we use the chemical model Nautilus in its 3-phase version from (Ruaud *et al.* 2016). The Nautilus code computes the gas-phase and dust ice composition as a function of time taking into account reactions in the gas-phase, sticking on grains and desorption from grain surfaces, and reactions at the surface. The ice is decomposed into a surface (the two most external monolayers of molecules) and a bulk. Both the surface and the bulk are chemically active based on the Langmuir-Hinshelwood mechanism with the formalism of Hasegawa *et al.* (1992). In the bulk though the diffusion of species is much less efficient as compared to the surface. All the details and surface parameters for diffusion are the same as in Ruaud et al. (2016). For the desorption from the surface (only), we consider thermal desorption and desorption induced by cosmic-rays (Hasegawa & Herbst 1993) as well as by exothermic chemical reactions (exothermicity of surface chemical reactions allows for the species to be desorbed after their formation) (Garrod *et al.* 2007). The Garrod et al. (2007) chemical desorption mechanism leads to approximately that 1% of the newly formed species desorb and 99% remain on the grain surfaces (this correspond to a factor of 0.01 in (Garrod et al. 2007)). The binding energies of the species to the surface have been updated from Wakelam et al. (submitted to Molecular Astrophysics) and the atomic oxygen is allowed to diffuse by tunneling (see Wakelam et al. submitted for discussion on this point).

---

[3] See http://spec.jpl.nasa.gov

The gas-phase network is based on kida.uva.2014[4] (Wakelam et al. 2015b)), with the modifications described in section 3. The surface network and parameters are the same as in (Ruaud *et al.* 2015) with some updates from Wakelam et al. (submitted). Following (Hincelin *et al.* 2015), the encounter desorption mechanism is included in the code. This mechanism accounts for the fact that the $H_2$ binding energy on itself is much smaller than on water ices and prevents the formation of several $H_2$ monolayers on grain surfaces.

The chemical composition of the gas-phase and the grain surfaces is computed as a function of time. The gas and dust temperatures are equal to 10 K, the total $H_2$ density is equal to $3\times10^4$ cm$^{-3}$ (various runs have been performed with total H density been varied between $1\times10^4$ cm$^{-3}$ and $2\times10^5$ cm$^{-3}$). The cosmic-ray ionization rate is equal to $1.3\times10^{-17}$ s$^{-1}$ and the total visual extinction is set equal to 10. All elements are assumed to be initially in atomic form (elements with an ionization potential below the maximum energy of ambient UV photons (13.6 eV, the ionization energy of H atoms) are initially in a singly ionized state, i.e., C, S, Si, Fe, Na, Mg, Cl, and P), except for hydrogen, which is entirely molecular. The initial abundances are similar to those of Table 1 of Hincelin *et al.* (2011), the C/O elemental ratio being equal to 0.7 in this study. The grains are considered to be spherical with a 0.1μm radius, a 3 g.cm$^{-3}$ density and about $10^6$ surface sites, all chemically active. The dust to gas mass ratio is set to 0.01.

[4] http://kida.obs.u-bordeaux1.fr/models

### 3.2. Update of the chemistry

To determine the unknown rate constant which will be used in the network, we use a methodology developed in previous articles (Loison et al. 2014a, Loison *et al.* 2014b, Loison *et al.* 2015) and summarized in appendix A. This methodology includes extensive literature review, various DFT and ab-initio calculations for critical gas phase reactions, namely H + l-$C_3H_2$, H + t-$C_3H_2$, O + c-$C_3H_2$, O + l-$C_3H_2$, N + $C_3$, N + c-$C_3H_2$, N + l-$C_3H_2$, H + $C_3O$, O + $C_3O$, O + c-$C_3H_3^+$, OH + $C_3$, OH + c-$C_3H_2$, $H_2$ + l-$C_3H$ and $H_2$ + c-$C_3H$, to determine the presence, or not, of barrier. When there is no barrier in the entrance valley and exothermic bimolecular exit channels, we choose to use capture rate constant or sometimes a fraction of the capture rate constant by comparison with similar reactions (capture rate is the upper limit of the rate constants for barrierless reactions). Dissociative Recombination (DR) of c,l-$C_3H_2^+$ and c,l-$C_3H_3^+$ is an important source of c,l-$C_3H$ and is the main source c,l,t-$C_3H_2$ in our network. The first step of c,l-$C_3H_2^+$ and c,l-$C_3H_3^+$ DR is the formation of highly excited $C_3H_2^{**}$ and $C_3H_3^{**}$, which leads to bond fragmentation. Angelova *et al.* (2004) have shown that the DR of c-$C_3H_2^+$ leads to 87.5 % of $C_3H_x$ and 12.5 % of $C_2H_y$ + $CH_z$, and the DR of c-$C_3H_3^+$ leads to 90.7% of $C_3H_x$ and 9.3% of $C_2H_y$ + $CH_z$. Moreover, in DR processes, the H ejection is in general favored than $H_2$ ejection (Plessis *et al.* 2012, Plessis *et al.* 2010, Janev & Reiter 2004). Considering the exothermicty for the ejection of two hydrogen atoms (endothermic for c-$C_3H_3^+$ DR and only slightly exothermic for l-$C_3H_2^+$, c-$C_3H_2^+$ and l-$C_3H_3^+$ DR, see annex B) this process will have low branching ratio. Then, dissociation of $C_3H_2^{**}$ and $C_3H_3^{**}$ will mainly produce $C_3H$ + H and $C_3H_2$ + H, both $C_3H$ and $C_3H_2$ species being also excited considering the exothermicity of the DR and the fact that hydrogen atom will carry only a limited part of the available energy through kinetic energy. Part of the excited $C_3H$ and $C_3H_2$ will lead to dissociation when they are populated above the dissociation limit, but most of them will relax through radiative emission of an infrared photon. As noted by Herbst et al (2000), the typical time-scales for isomeric conversion is much shorter than for relaxation by one infrared photon. Thus, as radiative relaxation occurs slowly, isomeric conversion leads to equilibrated isomeric (c-$C_3H$ ⇆ l-$C_3H$, c-$C_3H_2$ ⇆ l-$C_3H_2$) abundances at each internal energy. The final balance is determined at or near the effective barrier to isomerization, which corresponds to the energy of the transition state. The ratio between the isomeric forms are then **approximated** by the ratio of the rovibrational densities of states of the isomers at the barrier to isomerization calculated using MESMER program (Glowacki *et al.* 2012). Figure 2 shows the isomerization pathway calculated at DFT level. The calculated geometries of the stationary point can be found in appendix A. The t-$C_3H_2$ has a **triplet ground state** and its production in excited singlet ground state is neglected here. The production of t-$C_3H_2$ from DR of c,l-$C_3H_3^+$ is supposed to come from the c,l-$C_3H_3^+$ + $e^-$ → c,l-$C_3H_3$ → t-$C_3H_2$ + H which is supposed to be a minor channel versus c,l-$C_3H_3^+$ + $e^-$ → c,l-$C_3H_3$ → c,l-$C_3H_2$ + H (t-$C_3H_2$ has a ground triplet state contrary to c,l-$C_3H_2$ which have a ground singlet state).

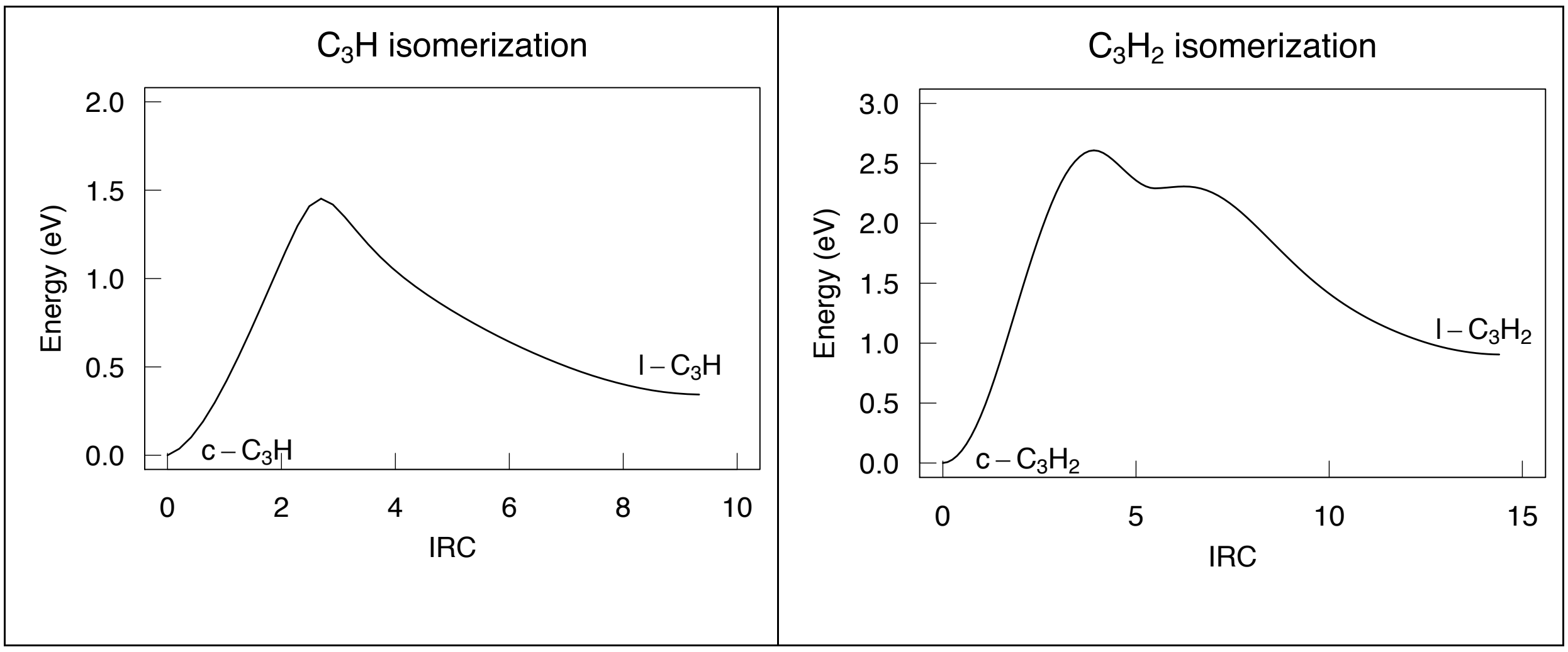

**Figure 2**: Intrinsic Reaction Coordinate pathway for the isomerization of c,l-$C_3H$ and c,l-$C_3H_2$ in their ground state calculated at M06-2X/aug-cc-pVTZ level.

Then, contrary to usual astrochemical models where it is assumed that the cyclic ions leads to cyclic neutral, and linear ions leads only to linear neutral, we consider that c,l-$C_3H_2^+$ DR leads mainly to cyclic c-$C_3H$ and c,l-$C_3H_3^+$ DR leads mainly to cyclic c-$C_3H_2$, with, in both cases, cyclic to linear branching ratio given by the rovibrational densities of states of the isomers at the barrier to isomerization. The reactions involving $C_3H_x$ chemistry, including c,l-$C_3H_2^+$ and c,l-$C_3H_3^+$ DR, are represented displayed Figure 3. It should be noted that if c,l,t-$C_3H_2$ are produced almost exclusively from c,l-$C_3H_3^+$ DR, the c,l-$C_3H$ are produced not only from c,l-$C_3H_2^+$ and c,l-$C_3H_3^+$ DR, but also through the C + $C_2H_2$ reaction as well as H + $C_3^-$ reaction.

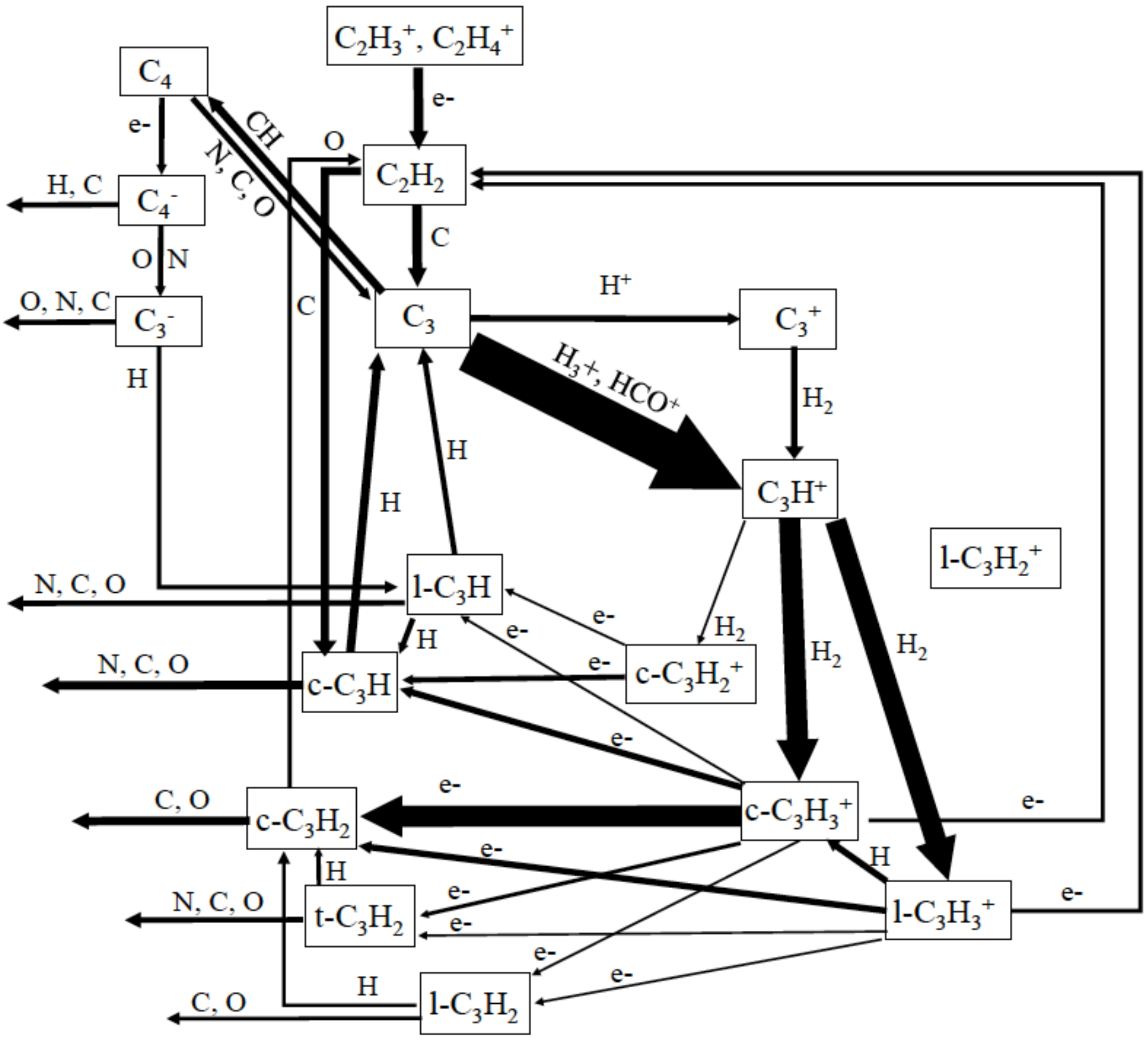

**Figure 3**. Schematic diagram highlighting the important pathways for $C_3H_x$ chemistry in dense molecular cloud for a typical age (between $1\times10^5$ and $6\times10^5$ years). The thickness of each arrow is proportional to the fluxes (the very minor fluxes are not represented).

## 4 Results and comparison with observations

With our network and the physical conditions relevant to the various molecular clouds, the main formation pathways for c,l,t-$C_3H_2$ is through DR of c,l-$C_3H_3^+$, these ions being formed mainly through $C_3$ + $H_3^+$, $HCO^+$, $HCNH^+$ → $C_3H^+$ reactions followed by $C_3H^+$ + $H_2$ reaction, the $C_3H^+$ being also produced through the $C^+$ + $C_2H_2$ reaction, and the main consumption is with atoms: O, C for the four isomers, H for c,l-$C_3H$ and l,t-$C_3H_2$ and N for c,l-$C_3H$ and t-$C_3H_2$, most of these reactions are introduced for the first time in astrochemical network thanks to new ab-initio calculations. For a typical dense cloud age (time between $10^5$ years and $10^6$ years), the c-$C_3H_2^+$ and c,l-$C_3H_3^+$ DR is an important source of c,l-$C_3H$, but the main c-$C_3H$ formation pathway is the C + $C_2H_2$ reaction. Considering the importance of c-$C_3H_2^+$ and c,l-$C_3H_3^+$ DR, the branching ratio of the $C_3H^+$ + $H_2$ reaction, particularly between c,l-$C_3H_3^+$ and c-$C_3H_2^+$ + H, may have some importance. The rate constant measurements at room temperature for the $C_3H^+$ + $H_2$ reaction clearly show a rate constant function of the pressure leading only to $C_3H_3^+$, the nature of the isomer being not clearly identified (Raksit & Bohme 1983, Smith & Adams 1987). A more recent experimental study (Savic & Gerlich 2005) leads to a high rate constant for the radiative association at low temperature with however a non-negligible branching ratio toward c-$C_3H_2^+$. A global theoretical analysis, including branching ratio calculation using RRKM theory, has been performed by (Maluendes *et al.* 1993b, Maluendes *et al.* 1993a) leading to the formation of both c,l-$C_3H_3^+$ isomers in roughly equal abundances. We adopt their rate constant expressions for these channels and we also consider some c-$C_3H_2^+$ production with a high uncertainty factor. To estimate the importance of the uncertainties in these branching ratios we performed various runs considering only c-$C_3H_3^+$ or l-$C_3H_3^+$ production showing that the nature of the $C_3H_3^+$ isomer is not negligible (up to 20% for the cyclic-to-linear $C_3H_2$ ratio) but not critical, the key point being the relaxation of the $C_3H_3^{**}$ and $C_3H_2^*$ produced. It should be noted also that the low branching ratio for c-$C_3H_2^+$ + H channel considered in this study leads to a negligible importance for this channel.

The results of the simulation for the various species studied here and for typical molecular cloud conditions (n($H_2$) = $3\times10^4$ cm$^{-3}$, T = 10K) are shown Figure 4 as well as observations for TMC-1.

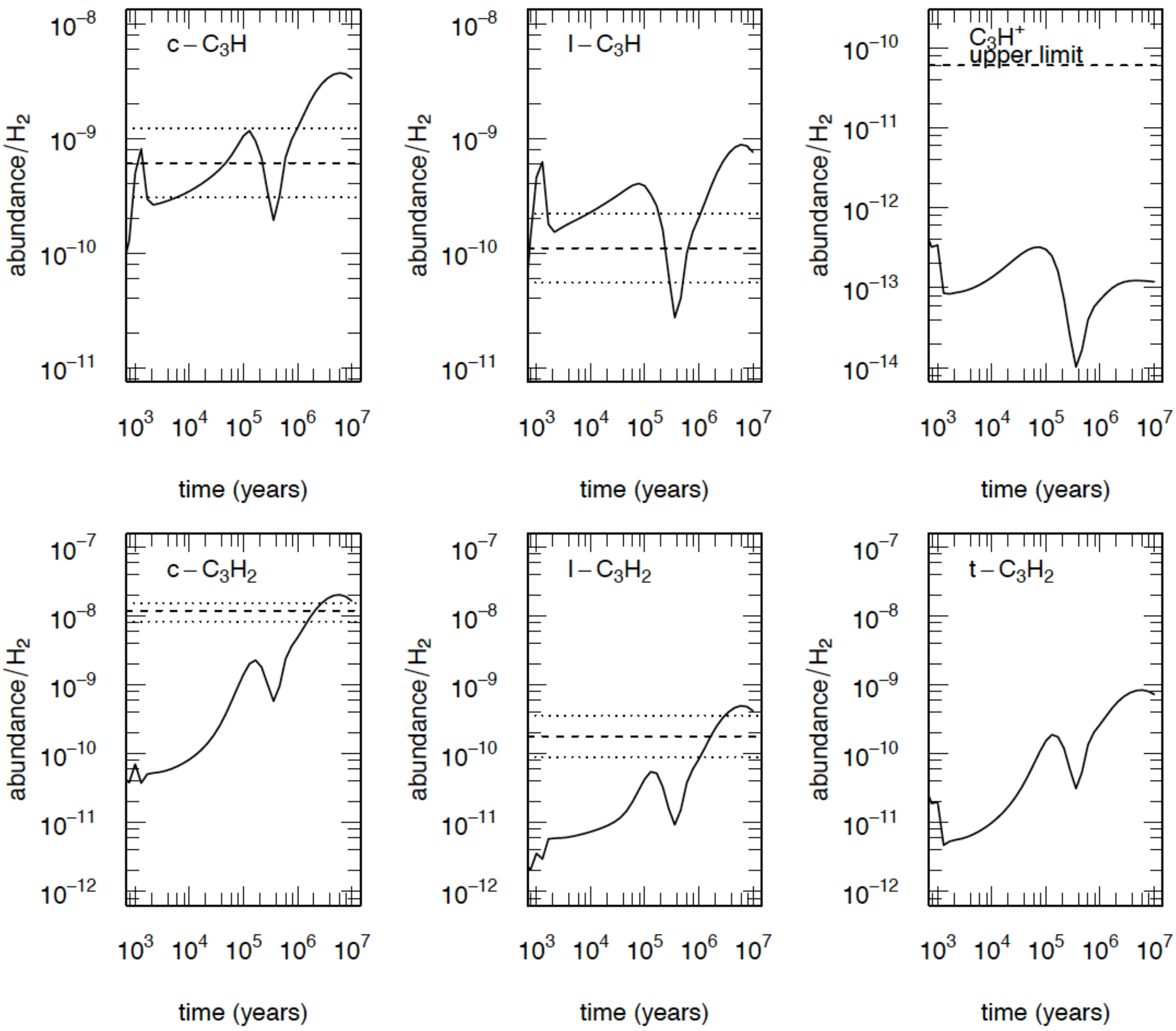

**Figure 4:** Abundances of $C_3H_x$ gas phase species studied in this work as a function of time predicted by our model ($n(H_2) = 3\times10^4$ $cm^{-3}$, T = 10K). The dashed horizontal lines represent the abundances observed for TMC-1 (this work, see Table 5). The dotted lines correspond to the uncertainties (see text).

**c-$C_3H$ and l-$C_3H$:**

The $C_3H$ isomers have been detected in this study only in two molecular clouds, TMC-1 and B1-b. The agreement between observations in TMC-1 and the model is good for a cloud time age between $10^5$ and $10^6$ years (see Figure 4), the age being not critical in that case. For TMC-1 and B1-b, the c-$C_3H$/l-$C_3H$ ratio is close to 5. It is worth noting that as shown on Figure 5, the c-$C_3H$/l-$C_3H$ ratio given by our model is almost independent of the total density of the molecular cloud in the range of $H_2$ between $1\times10^4$ and $2\times10^5$ molecule.$cm^{-3}$ in good agreement with observations (see table 5) (measured only for two molecular clouds however). In our model, c-$C_3H$ and l-$C_3H$ are produced not only from c,l-$C_3H_2^+$ and c,l-$C_3H_3^+$ DR, but also through the C + $C_2H_2$ reaction as well as H + $C_3^-$ reaction. The larger amount of cyclic isomer is due to the fact that both C + $C_2H_2$ reaction and DR of c,l-$C_3H_3^+$ and c,l-$C_3H_2$ favor c-$C_3$H. In the Horsehead PDR, the cyclic isomer is also favored with a slightly smaller ratio equal to 1.92 (Pety et al. 2012b). In IRC+10216 however, the linear isomer is more abundant with a c-$C_3$H/l-$C_3$H ratio equal to 0.38. In his simulation, **Agúndez (2009)** reproduces the fact that l-$C_3$H is more abundant than c-$C_3$H using the theoretical results from (Buonomo & Clary 2001) for the C + $C_2H_2$ reaction favoring l-$C_3$H production contrary to the more recent experimental results from (Costes et al. 2009). However, the result of (Costes et al. 2009) involves complex fitting leading to large uncertainty. Considering our chemical network and the photodissociation cross sections from van Hemert & van Dishoeck (2008), there is no alternative to the C + $C_2H_2$ reaction to form l-$C_3$H in circumstellar envelopes. Note that this reaction is the most important formation channel to l-$C_3$H in IRC+10216, but it is of little importance in molecular clouds. It can be noticed that c-$C_3$H is supposed to be mainly produced through c-$C_3H_2$ photodissociation in IRC+10216 (there is no experimental or theoretical data on c-$C_3H_2$ photodissociation branching ratio). There is then a crucial need to study the products of the photodissociations as well as for the C + $C_2H_2$ reaction despite the considerable amount of work already performed on it (Bergeat & Loison 2001, Buonomo & Clary 2001, Clary *et al.* 2002, Leonori *et al.* 2008, Costes et al. 2009, Hickson *et al.* 2016a).

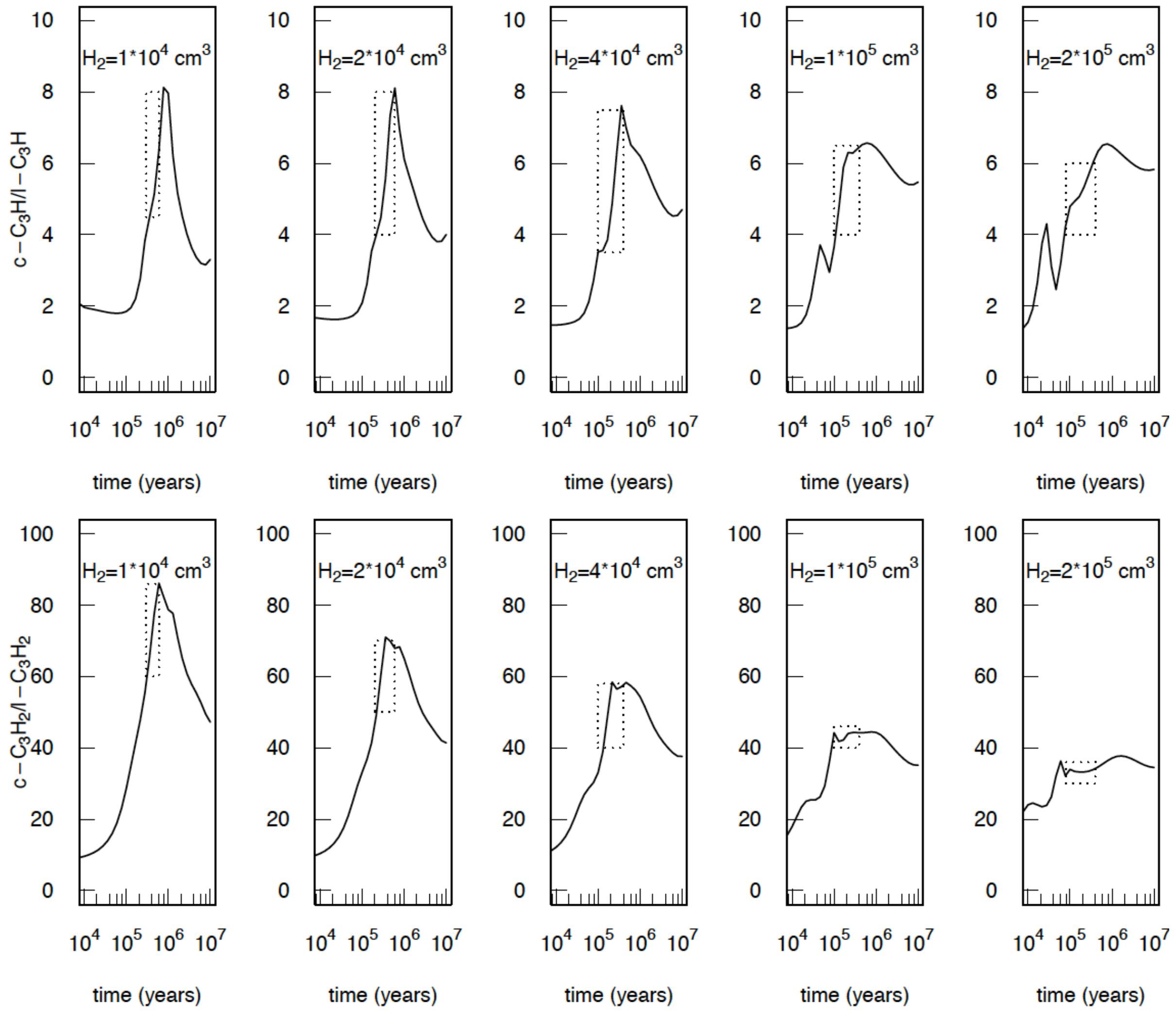

**Figure 5:** c-$C_3H$/l-$C_3H$ and c-$C_3H_2$/l-$C_3H_2$ ratios given by our model in function of the total density of the molecular cloud. Dotted rectangles correspond to the possible values at chemical cloud age given by the so-called distance of disagreement (Wakelam *et al.* 2006) at the given total densities.

**c-$C_3H_2$ and l-$C_3H_2$:**

The $C_3H_2$ isomers have been detected in this study in most of the molecular clouds. The agreement between observations in TMC-1 and the model is notably less good for c,l-$C_3H_2$ than for c,l-$C_3H$ for an early cloud age (few $10^5$ years). At late time (after $10^6$ years) the grain chemistry leads to relatively large s-$CH_4$ abundance (few %, compatible with observations on Ice in YSO (Boogert *et al.* 2015)). As $CH_4$ is not strongly bond to the ice (Raut *et al.* 2007), the release of $CH_4$ into the gas phase trigger an intensive hydrocarbon chemistry. Moreover, as for these clouds ages, oxygen and nitrogen atoms are depleted on grain leading mainly to $H_2O$, $CO_2$, $NH_3$ and HCN, all strongly bond to the ice with low desorption rate. Then, the hydrocarbons produced in the gas phase do not react with O and N atom but with C, others hydrocarbons neutral and ions leading to rich hydrocarbon chemistry and large $C_3H_x$ abundances. Among the various reactions, the C + $C_2H_2$ and $C^+$ + $C_2H_2$ reactions are of a particular importance. The most efficient c-$C_3H_2$ destruction pathway in dense molecular clouds is the O + c-$C_3H_2$ reaction which shows no barrier at various theoretical levels (DFT, MP2, CCSD(T), MRCI+Q detailed in Appendix A). Products and branching ratio for this reaction are not fully studied here but $HC_3O$ should be a non-negligible product of the O + c-$C_3H_2$ reaction.

The cyclic-to-linear $C_3H_2$ ratio observed here in 7 molecular clouds vary between 111 and 28 with decreasing values when the total density of the cloud increase. As the uncertainties for the observed cyclic-to-linear $C_3H_2$ ratios is estimated around 30%, these uncertainties cannot explain the difference between the various molecular clouds. In Figure 5, we have plotted the c-$C_3H_2$/l-$C_3H_2$ ratio evolution with the total density given by our model. The trend of the c-$C_3H_2$/l-$C_3H_2$ ratio is in good agreement with observations (see Table 5), for a chemical cloud age given by the so-called distance of disagreement (Wakelam et al. 2006) (minimum average difference (in magnitude) between modeled and observed abundances) represented with dotted rectangle on the Figure 5. As $C_3H_2$ isomers are mostly coming from DR of c,l-$C_3H_3^+$, DR independent of the total density of the molecular clouds, the main reaction impacting the cyclic to linear $C_3H_2$ ratio is supposed to be the l-$C_3H_2$ + H → c-$C_3H_2$ + H reaction. We have then performed a run suppressing the l-$C_3H_2$ + H → c-$C_3H_2$ + H reaction (presented in Appendix A). This suppression strongly reduce the cyclic-to-linear $C_3H_2$ ratio which become close to the ratio given by the cyclic-to-linear $C_3H_2$ ratio of the c,l-$C_3H_3^+$ DR and also close to the cyclic-to-linear $C_3H_2$ ratio of the most dense molecular clouds. This clearly shows that the high value of the cyclic-to-linear $C_3H_2$ ratio of low density molecular clouds such as TMC-1 and L1495B (see Table 5) is due to the l-$C_3H_2$ + H → c-$C_3H_2$ + H reaction which convert efficiently the linear l-$C_3H_2$ into the cyclic c-$C_3H_2$. This reaction becomes less efficient when the total density increase because the H atoms density is almost constant with the total density of the cloud contrary to the fluxes of the $C_3H_x$ and $C_3H_x^+$ chemistry (for example the $C_3$ + $H_3^+$ flux increase with the total density for typical molecular cloud age).

In the Horsehead Nebula PDR, the cyclic isomer is also favored but with a smaller ratio equal to 3.4 (Pety et al. 2012b). In IRC+10216, the cyclic-to-linear $C_3H_2$ ratio is ≈19 (Agúndez 2009), i.e., close to the values found in dense molecular clouds. It seems clear that, contrary to $C_3H$, the chemistry is comparable in all these objects, the formation of $C_3H_2$ isomers being mainly due to DR of $C_3H_3^+$, the photodissociation of $C_3H_3$ and $CH_3CCH$ playing a minor role. In PDR regions, the lower cyclic-to-linear $C_3H_2$ ratio may be due to the fact that c-$C_3H_2$ is supposed to be notably more photodissociated than l-$C_3H_2$ (van Hemert & van Dishoeck 2008). In diffuse clouds, where atomic H can be significant, the H + l-$C_3H_2$ $\rightarrow$ H + c-$C_3H_2$ reaction (introduced in this study) may strongly enhance the cyclic-to-linear $C_3H_2$ ratio and compensate the photodissociation effects.

**t-$C_3H_2$:**

We have introduced the third isomer of $C_3H_2$: t-$C_3H_2$ (HCCCH). This isomer, almost symmetric, has a stability between the cyclic and the linear isomer. It has a triplet ground state and it is reactive with H, C, O and also with N atom (cyclic and the linear isomer show barrier for reaction with N atoms). It should be noted that there is no measurement of the dipole moment of t-$C_3H_2$ and that the structure is not well defined, the $C_2$, $C_s$ and $C_{2v}$ structures being very close in energy (Aguilera-Iparraguirre et al. 2008). The dipole moment calculated by Nguyen et al. (2001) and equal to 0.51 D corresponds to the $C_2$ structure compatible with EPR spectroscopy (Seburg *et al.* 2009). Its introduction doesn't change notably the $C_3H$ and $C_3H_2$ chemistry but the reaction with N atom is a new pathway toward $HC_3N$ production, minor but non-negligible.

**$C_3H^+$:**

$C_3H^+$ has been detected only in three objects: the Orion Bar PDR, the Horsehead PDR, and Sgr B2(N) (Pety et al. 2012b, McGuire et al. 2013, McGuire et al. 2015). The lack of detection of l-$C_3H^+$ in cold dense molecular clouds (upper limit of $6\times10^{-11}\times[H_2]$ in TMC-1 from (McGuire et al. 2013)) is in agreement with our calculations where l-$C_3H^+$ abundance is around few $10^{-13}$ to $10^{-12}$ relative to $H_2$ for typical dense molecular cloud ages (few $10^5$ years). The formation of l-$C_3H^+$ is relatively well known through the reaction of $C^+$ with $C_2H_2$ and the reaction of $C_3$ with $H_3^+$, $HCO^+$ and $HCNH^+$. However, its main destruction rate via reaction with molecular hydrogen is not so well known despite the fact that the rate constant has been measured several times at various temperatures and pressures (Savic & Gerlich 2005, Raksit & Bohme 1983, Smith & Adams 1987). We use the global analysis performed by (Maluendes et al. 1993b, Maluendes et al. 1993a) leading to high rate constant at low temperature, which strongly limit the l-$C_3H^+$ abundance but has no, or very small, effect on c,l-$C_3H_3^+$ abundance and then almost no effect on c,l-$C_3H$ and c,l-$C_3H_2$, as long as the l-$C_3H^+$ consumption is dominated by reaction with $H_2$ instead of DR. The amount of l-$C_3H^+$ in the gas phase can be notably higher than the one given in our model if l-$C_3H^+$ reacts little bit more slowly with $H_2$ at low temperature which will have no, or small, effect on c,l-$C_3H$ and c,l-$C_3H_2$ abundances.

**l-$C_3H_3^+$, c-$C_3H_2^+$:**

Both l-$C_3H_3^+$ and c-$C_3H_2^+$ reach a non-negligible abundance in our calculation for typical dense molecular clouds (around few $10^{-11}$ and up to $10^{-10}$ versus $H_2$) and both have non-zero dipole moments (see Table 1). As already pointed out by Huang & Lee (2011) for l-$C_3H_3^+$, it may be detectable. However, it should be noted that the main formation pathway is the $C_3H^+$ + $H_2$ reaction for which the branching ratios toward l-$C_3H_3^+$ and c-$C_3H_2^+$ are not well known.

**The $C_3$ + O reaction**

As already highlighted in (Hickson et al. 2016b), $C_3$ is abundant in the gas phase as a result of various efficient neutral pathways producing $C_3$ and very few destruction mechanisms. The low reactivity of $C_3$ allows it to reach high abundance levels, which is at the origin of the rich $C_3H_4$, $C_3H_6$ and $C_3H_8$ production on grains through H addition reactions. To evaluate the role of the O + $C_3$ reaction, which has been studied only theoretically by (Woon & Herbst 1996) finding a small barrier in the entrance valley, we performed a run considering no barrier for this reaction and a rate constant equal to $2.0\times10^{-10}$ $cm^3$ $s^{-1}$ $molecule^{-1}$ (close to the capture rate constant) which considerably reduce the $C_3$ abundance. The introduction of O + $C_3$ reaction decreases all $C_3H_x$ abundances the agreement being notably less good, for typical dense cloud age, than considering a barrier for the O + $C_3$ reaction particularly for l-$C_3H_2$ and c-$C_3H_2$. At late time (after $10^6$ years) the release of $CH_4$, produced on grains, into the gas phase trigger an intensive hydrocarbon chemistry allowing to reproduce most of the observations. Considering the large uncertainty of grain reactions (most of the reactions on grain surface are not well characterized) and the large variability of the results with the description of the physic of the grains (for example induced by the new three phases model introduced by Ruaud *et al.* (2016)), the good agreement between observations and calculations at late age, considering no barrier for the O + $C_3$ reaction, may be fortuitous. It clearly shows the need to perform experimental measurement of the O + $C_3$ rate constant which will be a way to constraint the $CH_4$ formation on grains and its release in gas phase.

**5. Conclusion**

We have detected both cyclic and isomers $C_3H$ and $C_3H_2$ towards several molecular clouds using the IRAM 30m telescope. We have introduced in our gas-grain chemical model NAUTILUS new branching ratio of DR deduced from statistical theory with a relative amount of c-$C_3H$/l-$C_3H$ and c-$C_3H_2$/l-$C_3H_2$ being proportional to the density of vibrational states of each isomer near the effective barrier to isomerization, and not, as assumed in usual astrochemical models, that the cyclic ions lead only to cyclic neutral, and linear ions lead only to linear neutral. We have also introduced the third isomer of $C_3H_2$, t-$C_3H_2$ which play a minor but non-negligible role due to specific reactivity. Our model allows to reproduce the observations for c-$C_3H$, l-$C_3H$ and c-$C_3H_2$/l-$C_3H_2$ ratio in dense molecular clouds despite a notable underestimation for c-$C_3H_2$ and l-$C_3H_2$. In particularly the cyclic-to-linear $C_3H_2$ ratio given by our model is in good agreement with observations of the 7 molecular clouds studied in this paper, with a ratio value decreasing when the total density of the cloud increase. This decrease is due to the l-$C_3H_2$ + H → c-$C_3H_2$ + H reaction which is important for low density molecular cloud.

We also highlight the role of the O + $C_3$ reaction which is likely to possess a substantial barrier, in good agreement with the calculations of Woon & Herbst (1996) leading to a high $C_3$ abundance in dense molecular clouds.

Another critical point is the branching ratio of the C+ $C_2H_2$ reaction, the amount of l-$C_3H$ produced being critical to reproduce the linear to cyclic ratio in IRC+10216.

This work was supported by the program "Physique et Chimie du Milieu Interstellaire" (PCMI) funded by CNRS and CNES. VW researches are funded by the ERC Starting Grant (3DICE, grant agreement 336474).

M.A., N.M. and J.C. thanks the ERC for support under grant ERC-2013-Syg-610256 NANOCOSMOS. They also thank Spanish MINECO for funding support under grants AYA2012-32032, and from the CONSOLIDER Ingenio program "ASTROMOL" CSD 2009-00038. IRAM is supported by INSU/CNRS (France), MPG (Germany), and IGN (Spain).

This work received financial support from the French Agence Nationale de la Recherche (ANR) under grant ANR-13-BS05-0008 (IMOLABS : Molecules interstellaires : spectroscopie et synthèse en laboratoire)

We also thank the anonymous reviewer for his useful comments to improve the manuscript, particularly the suggestion to study more in detail the relation between c-$C_3H_2$/l-$C_3H_2$ ratio and the total $H_2$ density of the molecular cloud.